\documentclass[letterpaper]{jpconf}
\usepackage{graphicx}
\usepackage{subfigure}
\usepackage{amsmath}
\usepackage{amssymb}

\begin{document}

\title{Quarkonia studies with early ATLAS data}

\author{Sue Cheatham 
on behalf of the ATLAS Collaboration}

\address{Department of Physics, Lancaster University, Lancaster,  LA1 4YB, UK.}

\begin{abstract}
Studies of the production rates and properties for various charmonium and bottomonium states will have a decisive role in the understanding of the ATLAS detector with early data, and will also give an insight into the QCD production models.

\end{abstract}

\section{Introduction}
Amongst the first physics results of the ATLAS detector at LHC \cite{DetPaper} will be an analysis of J/$\psi$ and $\Upsilon$ production.  Even at low luminosity ATLAS will  quickly start improving on current quarkonia measurements providing world leading results with around 10~pb$^{-1}$ of data.  The narrow quarkonia resonances  will also be used as tools for alignment and calibration of the trigger and tracking systems.

\section{First data commissioning}
The narrow quarkonia resonances  will be used to gain understanding of first physics performance measurements.  Trigger validation, performance and efficiency studies will take place, allowing optimisation of trigger parameters such as thresholds and prescale factors.
  
Mass shifts of J/$\psi$ in physics variables can detect effects in different parts of the detector.  Preselection cuts ensure that the decay kinematics of the J/$\psi$ are particularly favorable for magnetic field studies since the muons from J/$\psi$ have large boosts and small opening angle, so magnetic field effects are not smeared out.
Di-muons from $\Upsilon$ have large opening angles which allows more accurate vertexing, useful for alignment studies.

Both muons from the J/$\psi$ have generally low $p_T$.  These complement the muon samples from Z decays which have  $p_T$ above $20~GeV$.  We expect 1~pb$^{-1}$ ( $\sim$1 day) muon stream early ATLAS data, to yield around 17,000 J/$\psi$ and 20,000 $\Upsilon$(1S) before trigger.  In comparison we only expect  400 Z $\to \mu \mu$ events with 1~pb$^{-1}$ .

CDF and ${D0}$ extensively and successfully used quarkonia mass shifts to disentangle various detector effects but it took many years at the Tevatron to collect sufficient statistics across a range of variables to allow for proper correction and alignment leading to a detailed understanding of various detector effects. 

\begin{figure}[htbp]
\begin{center}
\includegraphics[width=12cm]{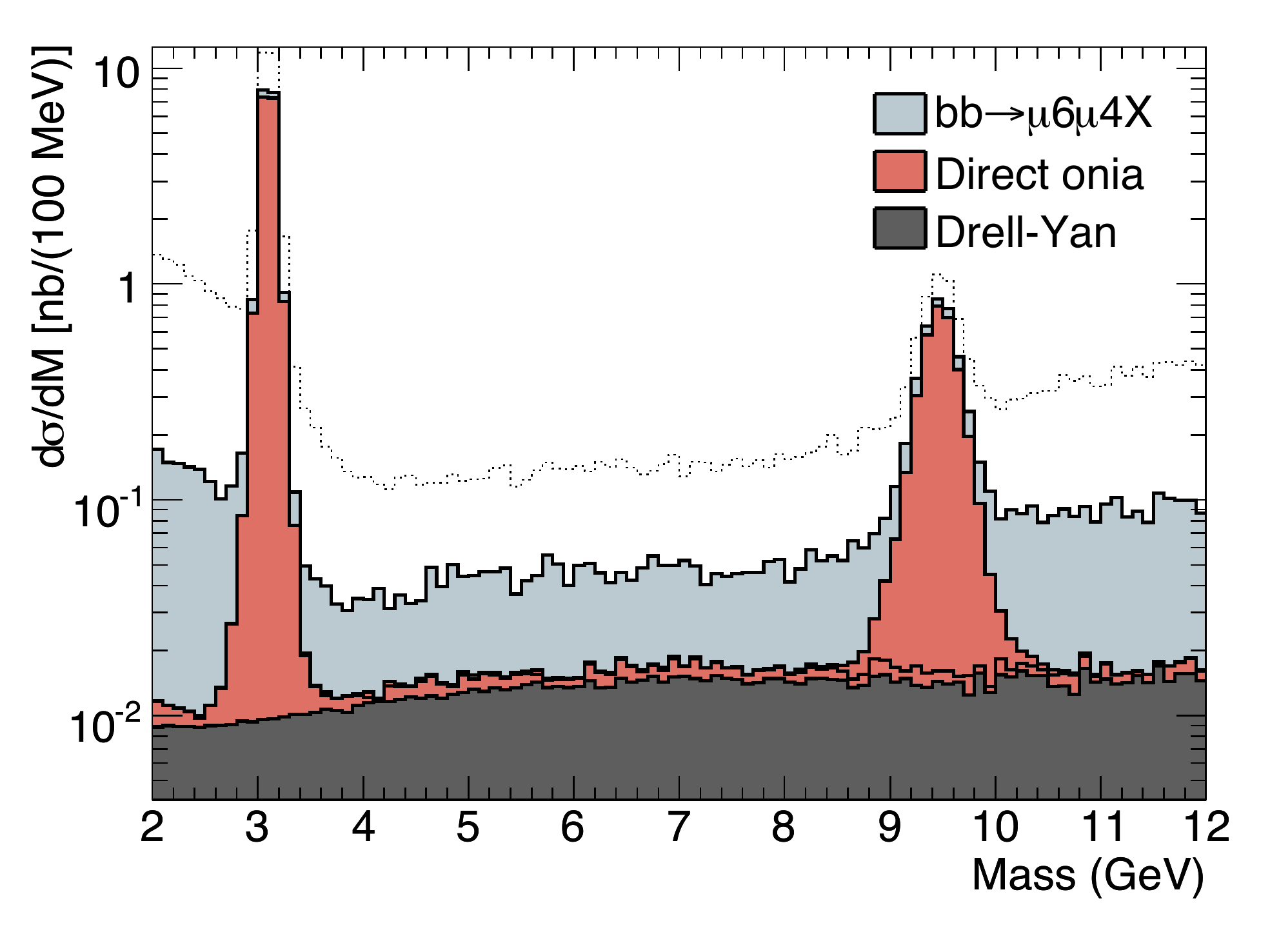}
\caption{Sources of low invariant mass di-muons, after background suppression cuts.  The dotted line shows the background level before vertexing and proper-time cuts.  \cite{OniaCSC} }
\label{fig:invMass}
\end{center}
\end{figure}

\section{Observation of prompt quarkonium production in ATLAS}
The dimuon mass resolutions at ATLAS are expected to be approximately $54~MeV$ (J/$\psi$) and $170~MeV$ ($\Upsilon$).  The main expected sources of background are indirect J/$\psi$ from B-decays, the continuum of muons from heavy flavour decays, Drell-Yan and decays in flight of  $K^\pm$ and $\pi^\pm$.  Figure~$\ref{fig:invMass}$ shows the reconstructed invariant mass distribution in the J/$\psi$ and $\Upsilon$ region for the di-muon dataset for an integrated luminosity of about 10~pb$^{-1}$.  The background is suppressed with vertexing and impact parameter cuts on the muons and a pseudo-proper time cut on the reconstructed quarkonium candidate.

\section{Separation of prompt and indirect J/$\psi$ }

The radial displacement, $L_{xy}$, of the two-track vertex from the beamline is used to distinguish between prompt and non-prompt J/$\psi$.  

Pseudo-proper time $t_0$ is defined by:

\begin{equation}
t_0 =  \frac{L_{xy}\cdot M_{J/\psi}}{p_T(J/\psi)\cdot c}
\end{equation}

\noindent where $M_{J/\psi}$ and $p_T(J/\psi)$ represent the J/$\psi$ invariant mass and
transverse momentum, $c$ is the speed of light in vacuum, and $L_{xy}$
is the measured transverse decay length (the displacement in the transverse $x-y$ plane
traversed by the J/$\psi$ from the primary vertex). 

\begin{figure}[htbp]
\begin{center}
\subfigure[Pseudo-proper time distribution]{
\includegraphics[width=6cm]{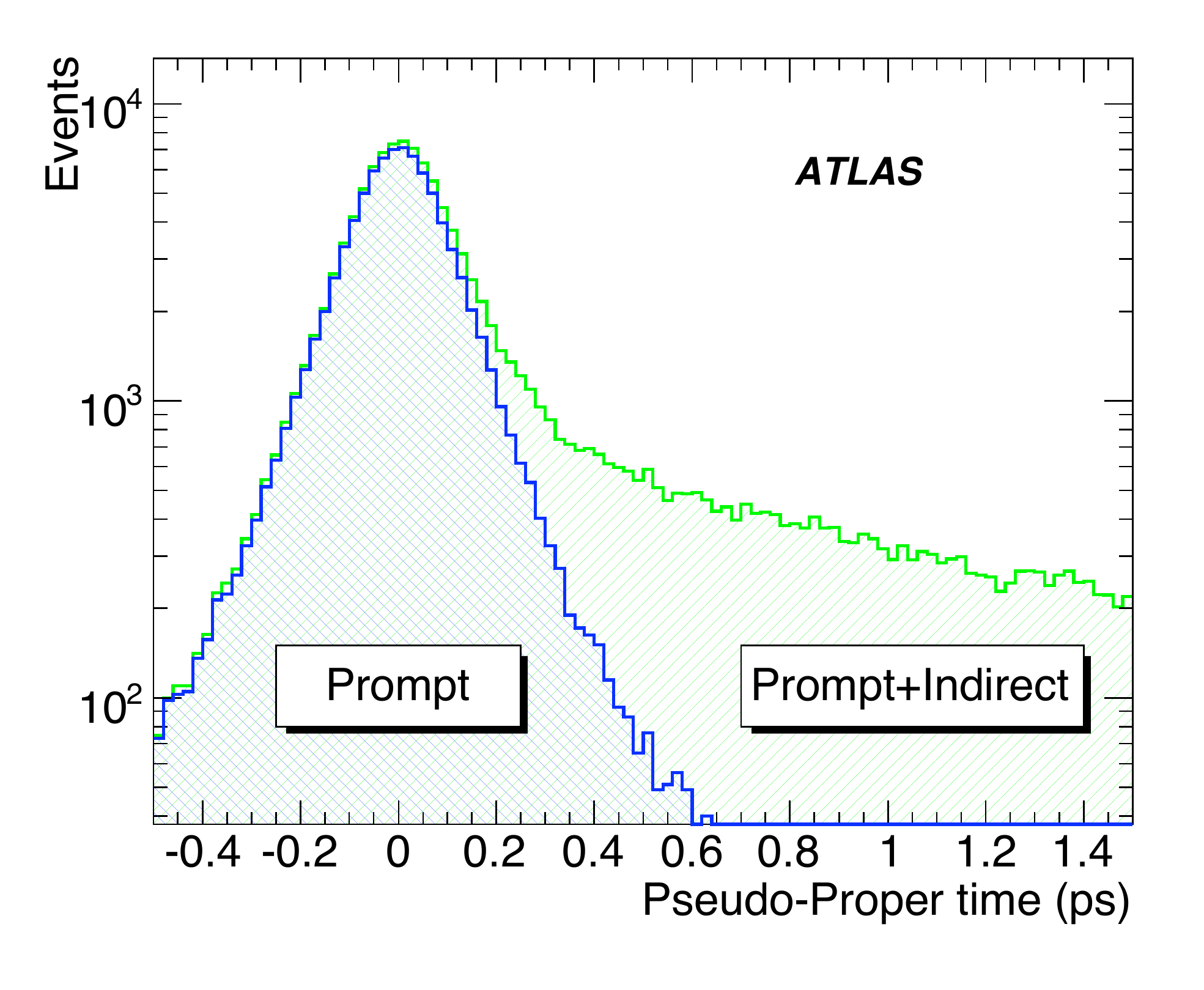}}
\subfigure[Efficiency and purity]{
\includegraphics[width=6cm]{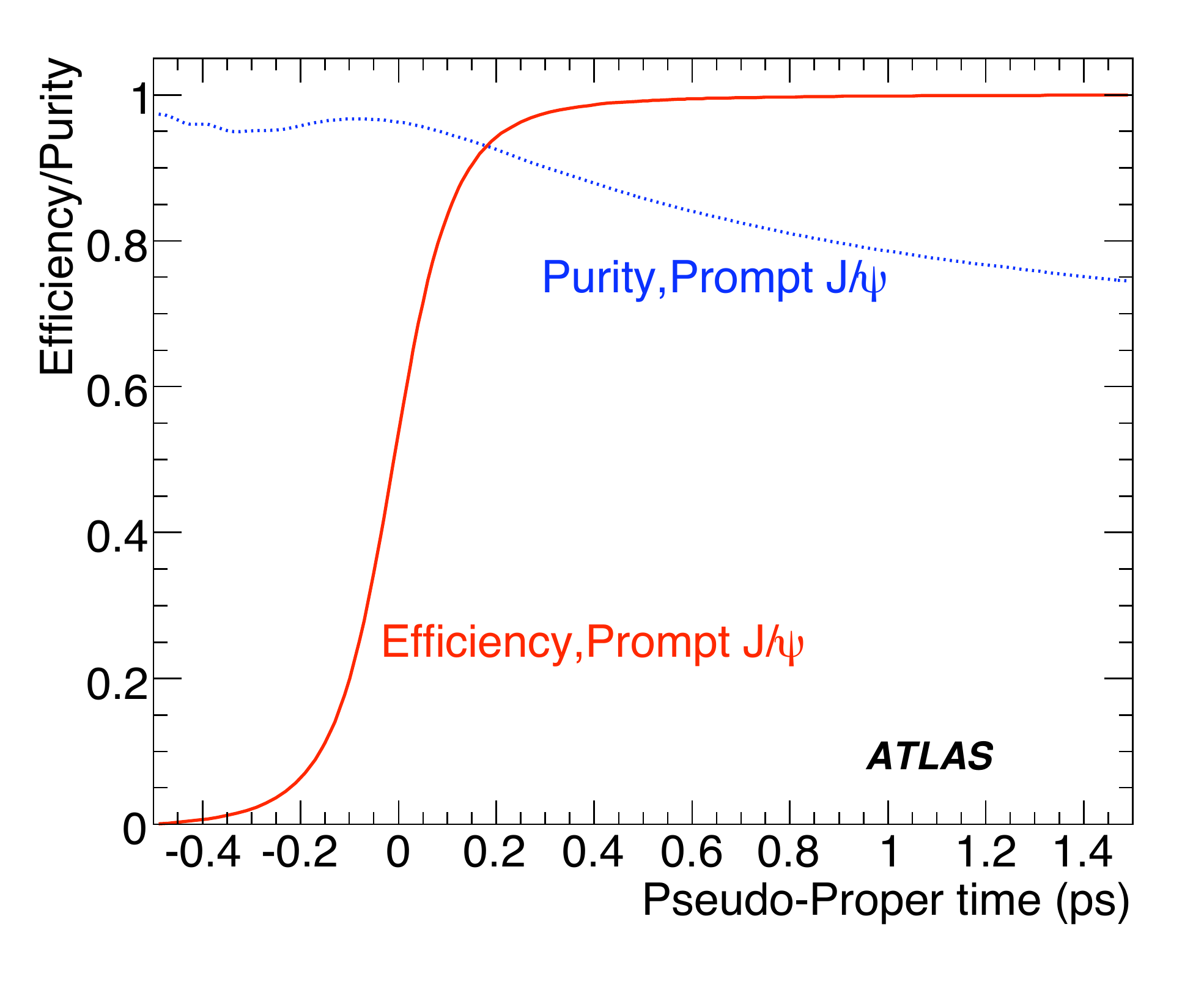}}
\caption{(a) Pseudo-proper time distribution for reconstructed prompt  J/$\psi$ and the indirect B-decay J/$\psi$ candidates.  (b) Efficiency (solid red line) and purity (dotted blue line) for prompt J/$\psi$ candidates as a function of the pseudo-proper time cut. \cite{OniaCSC} }
\label{fig:PromptInd}
\end{center}
\end{figure}

Prompt  J/$\psi$ have a zero pseudo-proper time whereas non-prompt J/$\psi$ primarily come from B-hadron decays with an exponentially decaying pseudo-proper time distribution due to the lifetime of the parent B-hadrons.
The distribution shown in Figure~$\ref{fig:PromptInd}$a is very powerful since a simultaneous fit gives access to a number of important measurements, such as the detector resolution in $t_0$, relative yields of prompt and non-prompt J/$\psi$, and effective average lifetime of B hadrons.

A pseudo-proper time cut of less than $0.2~ps$ allows one to retain prompt J/$\psi$ with an efficiency of 93\% and a purity of 97\% as shown in Figure~$\ref{fig:PromptInd}$b.

\section{Misalignment and material monitoring}
An important ongoing activity is the analysis of the effects of various types of deformations on the parameters of quarkonia resonances such as mass and pseudo-proper lifetime.  A key part of this effort is the development of a monitoring tool, capable of demonstrating the success (or otherwise) of the alignment procedure.

Another ongoing activity is the material mapping of the ATLAS Inner Detector.  Incorrect accounting of material can cause biases that can be detected as systematic shifts in the measured mean value of the J/$\psi$ mass.  In ATLAS we have developed a fast and efficient method for electron bremsstrahlung recovery, based on the dynamical adjustment of the ``system noise" term in the Kalman covariance matrix, once a bremsstrahlung like behavior has been detected.  This is called Dynamic Noise Adjustment (DNA).  The output of the Kalman DNA Fitter, position and intensity of the bremsstrahlung, can be used for material monitoring in the Inner Detector, since the probability of  bremsstrahlung increases with increased material thickness.

\section{Polarisation measurement}

Different QCD production models predict different spin alignment dependence with $p_T$.  There is little agreement between the models, and indeed between measurements, especially at low $p_T$.

The decay angle $\theta^\ast$ is defined as the angle between positive muon direction in the quarkonium rest frame and quarkonium direction in the lab frame.  Distribution of the decay angle depends on polarisation state.  Without considering spin alignment, the decay angle distribution is flat.  For J/$\psi$ with fully transverse polarisation,  spin alignment is described by $1+\cos^2~\theta^\ast$, whereas J/$\psi$ with fully longitudinal polarisation, spin alignment is described by $1-\cos^2~\theta^\ast$.  When the two muons forming a quarkonium state have similar $p_T$ the decay angle $\theta^\ast$ tends to be small.  As the decay angle opens, one muon tends to have a higher $p_T$, the other a lower  $p_T$.

\begin{figure}[htbp]
\begin{center}
\includegraphics[width=12cm]{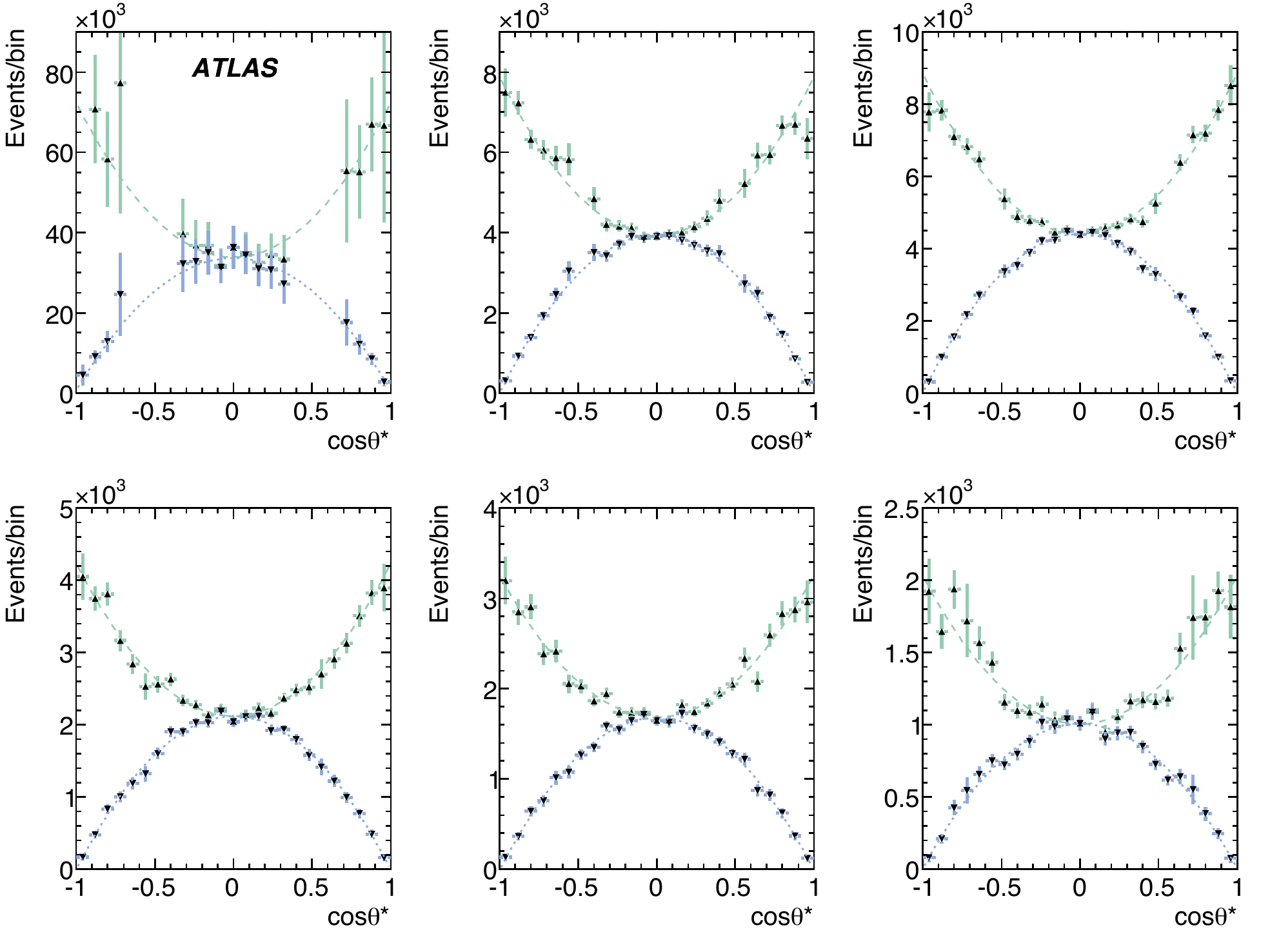}
\caption{Combined and corrected distributions in polarisation angle $\cos~\theta^\ast$, for longitudinally (dotted lines) and transversely (dashed lines) polarised J/$\psi$ mesons, in slices of J/$\psi$ $p_T$: left to right, top to bottom $9-12~GeV$, $12-13~GeV$, $13-15~GeV$, $15-17~GeV$, $17-21~GeV$, above $21~GeV$.  \cite{OniaCSC} }
\label{fig:Acc}
\end{center}
\end{figure}

In Tevatron studies of J/$\psi$ polarisation measurement a dimuon trigger was used, which limited the acceptance to the case where both muons have reasonably high $p_T$ and the decay angle is limited to the range $\pm~0.6$ .  ATLAS proposes to use not only a di-muon trigger, but also a single muon trigger with a 10~GeV threshold.  The single 10~GeV muon trigger will allow events to be picked up at large decay angle, $\theta^\ast$, combining the triggered muon with a reconstructed track ($p_T >$500~MeV) offline to form a quarkonium candidate with improved acceptance.

Areas with geometrical acceptances close to 100$\%$, as well as overlap measurements between the single and dimuon samples can be used for cross-normalisation of the two samples.

The di-muon and single muon samples are separately corrected for efficiencies and acceptances.
High $p_T$ data in both samples have 100$\%$ acceptance which is used for cross-normalisation and combination.
Figure~$\ref{fig:Acc}$ shows distribution of the reconstructed samples, corrected for acceptances and efficiencies, for two extreme polarisation scenarios.
Good agreement is achieved with simulated samples.

With an integrated luminosity of just 10~pb$^{-1}$, ATLAS aims to measure the polarisation of prompt vector quarkonium states to far higher transverse momenta than previous experiments, with extended coverage in decay angle.

\end{document}